\documentclass[onefignum,onetabnum, twoside]{article}

\usepackage{helvet}
\usepackage{dsfont}
\usepackage[T1]{fontenc}
\usepackage[utf8]{inputenc}
\usepackage{float}
\usepackage{pgf} 
\usepackage{booktabs}
\usepackage{amsmath}
\usepackage{amssymb}
\usepackage{graphicx}
\usepackage{microtype}
\usepackage{bbm}
\usepackage{xcolor}
\usepackage{adjustbox}
\usepackage{subcaption}
\usepackage{enumitem,kantlipsum}
\usepackage[space-before-unit, use-xspace]{siunitx}

\usepackage{array}
\usepackage{arydshln}
\makeatletter

\providecommand{\\}{\\}
\floatstyle{ruled}
\newfloat{algorithm}{tbp}{loa}
\providecommand{\algorithmname}{Algorithm}
\floatname{algorithm}{\protect\algorithmname}

\numberwithin{equation}{section}
\numberwithin{figure}{section}

\usepackage{leading}
\usepackage{comment}
\usepackage{ifthen}
\usepackage{xargs}

\def\x{{x}}

\newcommandx{\norm}[2][1=]{\ifthenelse{\equal{#1}{}}{\left\Vert #2 \right\Vert}{\left\Vert #2 \right\Vert^{#1}}}
\newcommandx{\normLigne}[2][1=]{\ifthenelse{\equal{#1}{}}{\Vert #2 \Vert}{\Vert #2\Vert^{#1}}}

\newcommandx{\functionspace}[2][1=+]{\mathbb{F}_{#1}(#2)}

\newcommandx{\VarDeux}[3][3=]{\operatorname{Var}^{#3}_{#1}\left\{#2 \right\}}

\newcommand{\LeftEqNo}{\let\veqno\@@leqno}

\newcommand{\N}{\ensuremath{\mathbb{N}}}

\newcommandx{\Vnorm}[2][1=V]{\| #2 \|_{#1}}
\newcommandx{\VnormEq}[2][1=V]{\left\| #2 \right\|_{#1}}

\newcommandx\probaMarkovTilde[2][2=]
{\ifthenelse{\equal{#2}{}}{{\widetilde{\mathbb{P}}_{#1}}}{\widetilde{\mathbb{P}}_{#1}\left[ #2\right]}}

\def\eqsp{\;}

\newcommandx{\weight}[2][2=n]{\omega_{#1,#2}^N}

\RequirePackage{xcolor}[2.11]
\definecolor{header1}{cmyk}{.9,.5,0,.35}
\definecolor{blue1}{cmyk}{.9,.7,0,0}
\definecolor{blue2}{cmyk}{.93,.95,.2,.07}
\definecolor{maroon}{cmyk}{.4,1,.3,.2}
\definecolor{gold1}{cmyk}{.2,.2,1,.1}
\definecolor{gray}{cmyk}{0,0,0,.5}
\definecolor{green1}{cmyk}{1,0,1,0}
\definecolor{proofcolor}{cmyk}{1,0,1,0}
\definecolor{red1}{cmyk}{0,1,.8,0}
\definecolor{orange1}{cmyk}{0,.55,1,0}
\definecolor{strip}{cmyk}{.6,.1,.1,.1}

\newcommandx\sequence[3][2=,3=]
{\ifthenelse{\equal{#3}{}}{\ensuremath{\{ #1_{#2}\}}}{\ensuremath{\{ #1_{#2}, \eqsp #2 \in #3 \}}}}
\newcommandx\sequenceD[3][2=,3=]
{\ifthenelse{\equal{#3}{}}{\ensuremath{\{ #1_{#2}\}}}{\ensuremath{( #1)_{ #2 \in #3} }}}

\newcommandx{\sequencen}[2][2=n\in\N]{\ensuremath{\{ #1_n, \eqsp #2 \}}}
\newcommandx\sequenceDouble[4][3=,4=]
{\ifthenelse{\equal{#3}{}}{\ensuremath{\{ (#1_{#3},#2_{#3}) \}}}{\ensuremath{\{  (#1_{#3},#2_{#3}), \eqsp #3 \in #4 \}}}}
\newcommandx{\sequencenDouble}[3][3=n\in\N]{\ensuremath{\{ (#1_{n},#2_{n}), \eqsp #3 \}}}

\newcommand{\opnorm}[1]{{\left\vert\kern-0.25ex\left\vert\kern-0.25ex\left\vert #1 
    \right\vert\kern-0.25ex\right\vert\kern-0.25ex\right\vert}}

\newcommandx{\CPE}[3][1=]{{\mathbb E}_{#1}\left[#2 \left \vert #3 \right. \right]} 
\newcommandx{\CPVar}[3][1=]{\mathrm{Var}^{#3}_{#1}\left\{ #2 \right\}}
\newcommand{\CPP}[3][]
{\ifthenelse{\equal{#1}{}}{{\mathbb P}\left(\left. #2 \, \right| #3 \right)}{{\mathbb P}_{#1}\left(\left. #2 \, \right | #3 \right)}}

\newcommandx{\osc}[2][1=]{\mathrm{osc}_{#1}(#2)}

\newcommand\coupling[2]{\Gamma(\mu,\nu)}

\renewcommand{\geq}{\geqslant}

\newcommandx{\KL}[2]{\text{KL}\left( #1 | #2 \right)}
\newcommandx{\KLbig}[2]{\text{KL}\left( #1 \middle| #2 \right)}

\leading{17.25pt}
\usepackage{adjustbox}
\usepackage{algorithmic}

\usepackage{import}
\usepackage{xifthen}
\usepackage{pdfpages}
\usepackage{transparent}

\makeatother

\usepackage{babel}

\usepackage[margin=1in]{geometry}
\usepackage{hyperref}
\usepackage{cleveref}
\usepackage{fancyhdr}

\newcommand{\funding}[1]{#1}

\newcommand{\headers}[2]{
  \fancyhead[LE]{#1}
  \fancyhead[RO]{#2}
}

\headers{Do Bayesian imaging methods report trustworthy probabilities?}{D. Y. W. Thong, C. K. Mbakam and M. Pereyra}

\title{Do Bayesian imaging methods report trustworthy probabilities?\thanks{Submitted to the editors DATE.
		\funding{This work was supported by the UK Research and Innovation (UKRI) Engineering and Physical Sciences Research Council (EPSRC) through grants EP/T007346/1 and EP/V006134/1.   }}}

\usepackage{authblk}

\author[1,2]{David Y. W. Thong \thanks{\texttt{david.yw.t+journals@gmail.com}}}
\author[1]{Charlesquin Kemajou Mbakam \thanks{\texttt{cmk2000@hw.ac.uk}}}
\author[1]{Marcelo Pereyra \thanks{\texttt{m.pereyra@hw.ac.uk}}}

\affil[1]{Maxwell Institute for Mathematical Sciences and Heriot-Watt University, Edinburgh, UK}
\affil[2]{KTH Royal Institute of Technology, Stockholm, Sweden}

\ifpdf
\hypersetup{
	pdftitle={Model assessment of Bayesian Imaging Models},
	pdfauthor={D. Y. W. Thong, Charlesquin Kemajou Mbakam and M. Pereyra}
}
\fi

\begin{document}

\maketitle

\begin{abstract}
Bayesian statistics is a cornerstone of imaging sciences, underpinning many and varied approaches from Markov random fields to score-based denoising diffusion models. In addition to powerful image estimation methods, the Bayesian paradigm also provides a framework for uncertainty quantification and for using image data as quantitative evidence. These probabilistic capabilities are important for the rigorous interpretation of experimental results and for robust interfacing of quantitative imaging pipelines with scientific and decision-making processes. However, are the probabilities delivered by existing Bayesian imaging methods meaningful under replication of an experiment, or are they only meaningful as subjective measures of belief? This paper presents a Monte Carlo method to explore this question. We then leverage the proposed Monte Carlo method and run a large experiment requiring 1,000 GPU-hours to probe the accuracy of five canonical Bayesian imaging methods that are representative of some of the main Bayesian imaging strategies from the past decades (a score-based denoising diffusion technique, a plug-and-play Langevin algorithm utilising a Lipschitz-regularised DnCNN denoiser, a Bayesian method with a dictionary-based prior trained subject to a log-concavity constraint, an empirical Bayesian method with a total-variation prior, and a hierarchical Bayesian Gibbs sampler based on a Gaussian Markov random field model). We find that, a few cases, the probabilities reported by modern Bayesian imaging techniques are in broad agreement with long-term averages as observed over a large number of replication of an experiment, but existing Bayesian imaging methods are generally not able to deliver reliable uncertainty quantification results.

\end{abstract}

\begin{keywords}
\sloppy Bayesian inference, inverse problems, image processing, mathematical imaging, uncertainty quantification, credible intervals, data-driven regularisation.
\end{keywords}

\begin{MSCcodes}
62F15, 62G05, 65J20, 65J22, 65R32, 65C05, 68U10, 52H35, 65C60, 94A08
\end{MSCcodes}

\section{Introduction}
\sloppy Digital images provide essential information for decision-making in fields with major economical, societal, and environmental impact (e.g., in medicine, agriculture, forestry, astronomy, and defense). Early stages of image-based decision making were predominantly qualitative. However, over the past decades these field have adopted more rigorous decision-making strategies that require using image data in a quantitative manner. An essential aspect of the quantitative imaging paradigm is to reliably quantify the uncertainty in the delivered solutions. This is especially important in imaging problems that are not well posed, and which consequently have a particularly high uncertainty in the solution. Uncertainty quantification (UQ) provides essential information for the rigorous interpretation of experimental results and for robust interfacing of imaging pipelines with decision-making processes \cite{Bardsley2018-wx,Sullivan2019-dr}.

While computational imaging problems can be formulated and solved with many different mathematical and computational frameworks \cite{kaipio2006statistical,chambolle2016introduction,pereyra2016survey,Ongie2020,Mukherjee23,chen2023equivariant}, UQ for computational imaging relies strongly on the Bayesian statistical framework \cite{kaipio2006statistical,robert2007bayesian}. Indeed, Bayesian statistical inference provides a particularly suitable underpinning for UQ because of its foundation in probability theory and its powerful mathematical decision theory \cite{robert2007bayesian}. Since the seminal works \cite{besag1986statistical,Geman1984} four decades ago, a range of Bayesian imaging strategies have been proposed in the literature, employing a rich variety of modelling and algorithmic approaches (see, e.g., \cite{kaipio2006statistical,pereyra2016survey,Mukherjee23}). The field has advanced greatly over this period, with established Bayesian imaging technique being regularly superseded by newer ones offering superior image estimation accuracy and computational efficiency. 

Unlike image restoration accuracy, the accuracy of the UQ results produced by Bayesian imaging methods remains largely unexplored. The reasons are manifold. First, there has been an emphasis on developing the algorithmic infrastructure required to compute UQ results, rather than on UQ-accurate models. Second, assessing the accuracy of imaging UQ results is difficult and highly computationally expensive, as it requires large-scale replication experiments. Also, the field is transitioning and UQ results are predominantly analysed visually and qualitatively.

This paper seeks to tackle this critical gap in the literature by exploring the following key question which underpins all Bayesian UQ tasks in the context of imaging sciences: 

\begin{center}
\vspace{0.1cm}
\emph{Are the probabilities delivered by Bayesian imaging methods meaningful under replication of an experiment, or are they only meaningful as subjective measures of belief?}
\vspace{0.1cm}
\end{center}
In other words, suppose a Bayesian imaging technique is used repeatedly to produce probabilistic statements about unknown images of interest (e.g., as part of an imaging pipeline). Can we expect these statements to be broadly in agreement with probability expressed as a long-run relative frequency observed over infinitely many experiments? Or are the reported probabilities to be interpreted only subjectively?

The remainder of the paper is organised as follows. \Cref{sec:Section 2} introduces notation and a formal problem statement related to the accuracy of UQ results in the context of imaging sciences. Following on form this, \Cref{sec:Section 3} presents a Monte Carlo methodology for evaluating the accuracy of probabilities reported by Bayesian imaging techniques. This methodology is then used in \Cref{sec:experiments} to evaluate some important classes Bayesian image models that are central to the modern Bayesian imaging literature, spanning from the conventional total-variation Markov random field prior, to more recent data-driven priors such as score-based denoising diffusion priors. Our findings suggest that assessing the quality of a Bayesian imaging methods exclusively through the accuracy of the reconstructed images can be misleading, as models can simultaneously produce accurate image reconstruction and poor uncertainty quantification. Conclusions and perspectives for future work are finally reported in \Cref{sec:conclusion}.

It is worth mentioning at this point that currently the most accurate imaging UQ results are achieved by using the frequentist bootstrap method of Tachella and Pereyra \cite{Tachella24}, which is not a Bayesian method. Bayesian methods only achieve comparable state-of-the-art performance on very small problems (MNIST) and by using generative machine learning priors \cite{holden2022bayesian}. However, the bootstrap method of \cite{Tachella24} cannot support form of UQ derived from Bayesian decision theory. We hope that the availability of methodology for evaluating the accuracy of probabilities reported by Bayesian imaging techniques will allow the development of better statistical image models capable of supporting many forms of advanced UQ.

\section{Problem statement}\label{sec:Section 2}
\subsection{Bayesian modelling}
We seek to evaluate Bayesian imaging techniques that perform inference on an unknown image $x^{*}$ taking values in some space of images $\mathbb{X}$, from some observed data $y \in \mathbb{Y}$. Rather than image estimation accuracy, we will assess the Bayesian techniques based on their suitability for supporting UQ tasks. More precisely, we seek to check if the probabilities delivered by the Bayesian techniques are useful to a practitioners that intends to apply the considered imaging technique repeatedly, as part of an imaging pipeline, and wish to relate the obtained probabilities to long-run relative frequencies observed over a very large number of experiments. Within this context, we pay special attention to imaging inverse problems that are ill-conditioned or ill-posed. Such problems have high intrinsic uncertainty, and  the choice of prior distribution has a strong impact on the delivered uncertainty estimates.

Rooted in probability theory, Bayesian imaging methods represent the unknown image $x^{*}$ as a realization of a random variable $\mathbbm{x}$ taking values in the space of images $\mathbb{X}$. The data $y$ is modelled as a realisation of the $\mathbb{Y}$-valued conditional random variable $(\mathbbm{y}|\mathbbm{x}=x^{*})$. Having observed the outcome $\mathbbm{y}=y$, to perform inference on $\mathbbm{x}$, we specify a likelihood function $x \mapsto p(y|x)$ and a prior density function $x \mapsto p(x)$, and use Bayes' theorem to derive the posterior distribution for $(\mathbbm{x}|\mathbbm{y}=y)$ \cite{robert2007bayesian}. The resulting posterior density is given by
\begin{equation}\label{posterior}
p(x|y)=\frac{p(y|x)p(x)}{\int_{\mathbb{X}} p(\tilde{x},y)\textrm{d}\tilde{x}}\, .    
\end{equation}

When using the model \eqref{posterior} for inference, it is useful to recall that the joint density $p(x,y) = p(y|x)p(x)$ underpinning \eqref{posterior} describes the imaging problem at a ``population'' level. For illustration, suppose that $x^{*}$ is an image of an adult human face. The random variable $\mathbbm{x}$, with density $p(x)$, provides a probabilistic description of what an adult human face ``looks like'' for a generic member of the considered population, as acquired by an ideal camera with no noise or resolution limitations. The image $x^{*}$ is a realization of $\mathbbm{x}$ stemming from imaging a specific individual from that population with that perfect camera. The data ${y}$ represents the actual measurement or observation of $x^{*}$ in realistic instrumental conditions involving degradation from noise and limited resolution, as modeled by $p(y|x)$. The posterior \eqref{posterior} results from combining the population descriptions of $(\mathbbm{x},\mathbbm{y})$ provided by $p(x,y) = p(y|x)p(x)$, together with the value $y$ observed. In this paper, we seek to assess if probabilistic statements derived from \eqref{posterior} are meaningful when this process is repeated over many individuals in this population.
                
Constructing $p(x|y) \propto p(y|x)p(x)$ for a given imaging problem involves some important modelling choices. Most Bayesian imaging techniques will implement broadly similar choices for the likelihood function $p(y|x)$. These are generally motivated by the physics underpinning the data acquisition process and are stated explicitly in the description of the method. Conversely, the choices pertaining to the prior $p(x)$ are many and varied, and they have evolved significantly as the field has matured. Until recently, image priors were predominantly assumption-driven and represented subjective prior knowledge about expected structural or regularity properties of the desired solution (e.g., smoothness, sparsity, piece-wise regularity, self-similarity - see, e.g., \cite{buades_review_2005, Orieux2010, roininen2019hyperpriors, chatterjee2012patch, pan2013l0}). But modern image priors are increasingly data-driven and use machine learning strategies to encode prior knowledge that is available as datasets \cite{Ongie2020,Mukherjee23, gonzalez2022solving,holden2022bayesian}. 

Progress in modelling for Bayesian imaging sciences has been largely driven by estimation accuracy and computational efficiency considerations. Indeed, following decades of progress in statistical image models, modern Bayesian imaging models for $(\mathbbm{x}|\mathbbm{y}=y)$ are often capable of delivering remarkably accurate point estimates of $x^\star$ in reasonably short computing times. In addition, modern methods often also offer detailed theoretical guarantees on the delivered solutions, such as convergence theory for algorithms that compute solutions iteratively \cite{salsa2010fast,Durmus2018,laumont_bayesian_2022,Mukherjee23}. However, the steady progress in statistical image models also provides an opportunity to engage with more advanced forms of statistical thinking and pursue complex inferences beyond point estimation for image reconstruction. For example, as mentioned previously, there is increasing interest in strategies for performing UQ on $(\mathbbm{x}|\mathbbm{y}=y)$, which is one of the main advantages of formulating and solving imaging problems in the Bayesian statistical framework. Accurate UQ for $(\mathbbm{x}|\mathbbm{y}=y)$ would allow interfacing imaging pipelines with subsequent scientific analyses robustly, and improve the value of the restored images as evidence for decision-making under uncertainty. 

However, Bayesian models are not accurate or inaccurate in absolute terms, they can reliably support some inferences and fail in others. In particular, some models for $(\mathbbm{x}|\mathbbm{y}=y)$ deliver remarkably accurate image restoration results, which depend mainly on the posterior mean or mode, but perform poorly in UQ tasks that rely on higher moments and tail probabilities, as we illustrate in \Cref{sec:experiments}. Hence the need for methods to probe the accuracy of UQ techniques.

\subsection{Evaluating UQ accuracy via credible regions}
For presentation clarity, and without loss of generality, we frame our evaluation of the accuracy of Bayesian UQ results around Bayesian confidence intervals, or so-called \emph{credible regions}. A credible region of level $\alpha \in (0,1)$ is a subset $\mathcal{C}^y_\alpha$ of the solution space $\mathbb{X}$ such that $\mathbbm{x}$ takes values in $\mathcal{C}^y_\alpha$ with posterior probability $(1-\alpha)$, i.e.,
$$
\textrm{P}[\mathbbm{x}\in\mathcal{C}^y_\alpha|\mathbbm{y}=y] = 1-\alpha\, .
$$
In lay terms, $\mathcal{C}^y_\alpha$ is the Bayesian model's educated guess of the value of the unknown image $x^\star$, when it is allowed report a set of highly likely solutions rather than collapsing the guess to a single point estimate (i.e., a Bayesian estimator) \cite{robert2007bayesian}.
The modelling choices embedded in the posterior $p(x|y)$ can greatly influence UQ results. As mentioned previously, these choices influence UQ estimates such as $\mathcal{C}^y_\alpha$ more and differently than estimators of $(\mathbbm{x}|\mathbbm{y}=y)$. Therefore, it is essential to evaluate both image restoration accuracy and the accuracy of $\mathcal{C}^y_\alpha$. The latter are useful on their own (e.g., for hypothesis testing \cite{repetti2019scalable, liaudat2023_3}), and as proxies for other UQ forms.

A main difficulty when evaluating the accuracy of Bayesian models and the associated UQ results is that these can only be evaluated extrinsically. That is, if $(\mathbbm{x},\mathbbm{y})$ were genuinely distributed with density $p(x,y)$ and the algorithm used to $\mathcal{C}^y_\alpha$ form $p(x|y)$ were exact, then the probability $\textrm{P}[\mathbbm{x}\in\mathcal{C}^y_\alpha|\mathbbm{y}=y] = 1-\alpha$ would hold exactly. Consequently, all quantities computed from a Bayesian model are by construction exact in a self-referential manner. Of course, the crux of statistical inference is that all statistical models are imperfect, so the probabilities reported by $p(x|y)$ and summarised by $\mathcal{C}^y_\alpha$ will not always be accurate proxies for probabilities of real-world events under replications of an experiment. In particular, the empirical probability that $\mathcal{C}^y_\alpha$ contains $x^\star$ given $\mathbbm{y}=y$ could significantly differ from $1-\alpha$ when this probability is measured by applying the Bayesian imaging technique under scrutiny repeatedly over a large number of experiments. Evaluating the accuracy of probabilities reported by Bayesian models requires external information; ideally, benchmarking against the true distribution of $(\mathbbm{x},\mathbbm{y})$. In practice, our proposed method will use an empirical approximation of the true distribution of $(\mathbbm{x},\mathbbm{y})$ provided by a test dataset.


\section{Proposed method}\label{sec:Section 3}
\subsection{Proposed error measure}
We are now ready to present our proposed methodology for evaluating if the probabilities reported by Bayesian imaging techniques are in agreement with long-run relative frequency observed over many trials. To construct our method, we adopt an \emph{M-complete} Bayesian modelling approach \cite{Bernardo1994} and explicitly assume that there exists a true, albeit unknown, marginal distribution for $\mathbbm{x}$. This is the distribution that \emph{nature} would use as prior distribution to perform Bayesian inference on $(\mathbbm{x}|\mathbbm{y}=y)$. For presentation clarity, we assume that this true prior distribution admits a density w.r.t. the Lebesgue measure on the ambient space and denote this density by $p^\star(x)$. Similarly, we use $p^\star(x|y)$ to denote the density of the associated posterior distribution for $(\mathbbm{x}|\mathbbm{y}=y)$; i.e., $p^\star(x|y) \propto p(y|x)p^\star(x)$. As mentioned previously, we also assume the availability of a test dataset $\{x_i\}_1^M$, which we regard as realizations of i.i.d. copies of $\mathbbm{x}$ and hence as representatives of the prior $p^\star(x)$.

Basing inferences on \emph{nature}'s true distributions leads to Bayesian probabilities that perfectly agree with a frequentist definition of probability, as a limit of the relative frequency observed over infinitely many repetitions of an experiment in which $x^\star$ and $y$ are drawn from the joint distribution of $(\mathbbm{x},\mathbbm{y})$. In particular, our method exploits the fact that any credible region $\mathcal{C}^{\star,y}_\alpha$ constructed with \emph{nature}'s posterior distribution, such that 
$$
\textrm{P}[\mathbbm{x}\in\mathcal{C}^{\star,y}_\alpha|\mathbbm{y}=y] = \int_{\mathcal{C}_\alpha^{\star,y}} p^\star(x|y) \textrm{d}x = 1-\alpha\,,
$$
also verifies
\begin{equation}\label{Exp_C_alpha_star}
\textrm{E}_\mathbbm{y}\{\textrm{P}[\mathbbm{x}\in\mathcal{C}^{\star,\mathbbm{y}}_\alpha|\mathbbm{y}]\} = 1-\alpha\, ,
\end{equation}
for all $\alpha \in (0,1)$, where we note that in \eqref{Exp_C_alpha_star} the expectation is taken w.r.t. to marginal distribution of $\mathbbm{y}$. This results follow directly from the fact that, when operating with the true distribution for $(\mathbbm{x},\mathbbm{y})$, Bayesian inference is a straightforward application of probability theory, so all probability statements are perfectly calibrated. 

A closer analysis of the expectation \eqref{Exp_C_alpha_star} indicates that if we draw a realization $x^\star$ from $\mathbbm{x}$, then generate a measurement $y$ by drawing from the conditional r.v. $(\mathbbm{y}|\mathbbm{x}=x^\star)$ associated with the forward observation model, and subsequently use the true posterior density $p^\star(x|y) \propto p(y|x)p^\star(x)$ to identify a region $\mathcal{C}^{\star,y}_\alpha$ that accumulates $1-\alpha$ posterior probability mass, then the original realization $x^\star$ falls within that region with probability $1-\alpha$ when this probability is measured over replications of the experiment. This holds for all $\alpha \in [0,1]$ and all credible regions. Therefore, for these perfect true models, the quantification of the uncertainty is exact.

Of course, the oracle model $p^\star(x|y)$ and its associated region $\mathcal{C}^{\star,\mathbbm{y}}_\alpha$ are not available in practice, which is the crux of Bayesian modeling and the reason why external frequentist evaluation of Bayesian procedures is relevant. Indeed, if we repeat the above thought experiment but replace $p^\star(x|y)$ with any other Bayesian imaging model $p(x|y)$ to identify a region of the solution space $\mathcal{C}^{\mathbbm{y}}_\alpha$ that accumulates $1-\alpha$ probability, then the original realization $x^\star$ will not generally fall within that region with probability $1-\alpha$ over replications of the experiment. Our method seeks to quantify this mismatch in reported Bayesian probabilities versus observed empirical probabilities. This provides a proxy for assessing the impact of operating with the imperfect model $p(x|y)$, instead of $p^\star(x|y)$, when computing UQ results.

More precisely, with UQ applications in mind, and complementary to other notions of model quality, we propose to characterize the accuracy of $p(x|y)$ as a model for $(\mathbbm{x}|\mathbbm{y}=y)$ by benchmarking the set of HPD regions $\mathcal{C}^{\mathbbm{y}}_\alpha$ derived from $p(x|y)$, against the set of true regions $\mathcal{C}^{\star,\mathbbm{y}}_\alpha$, for a range of relevant values of $\alpha \in [0,1]$. In order to achieve this, we compute the expected difference in probability mass between $\mathcal{C}^{\mathbbm{y}}_\alpha$ and $\mathcal{C}^{\star,\mathbbm{y}}_\alpha$ across replications of the experiment, averaging over \emph{nature}'s joint distribution of $(\mathbbm{x},\mathbbm{y})$. In its simplest form, this leads to the following signed error measure:
\begin{equation}\label{loss}
\begin{split}
\ell(\mathcal{C}^{\mathbbm{y}}_\alpha, \mathcal{C}^{\star,\mathbbm{y}}_\alpha) &\triangleq \int [\boldsymbol{1}_{\mathcal{C}^{{y}}_\alpha}(x) -\boldsymbol{1}_{\mathcal{C}^{\star,{y}}_\alpha}(x)] \,p^\star(x)p(y|x) \textrm{d}x \textrm{d}y\, ,\\
&=\textrm{E}_\mathbbm{y}\{\textrm{P}[\mathbbm{x}\in\mathcal{C}^{\mathbbm{y}}_\alpha|\mathbbm{y}]\} -\textrm{E}_\mathbbm{y}\{\textrm{P}[\mathbbm{x}\in\mathcal{C}^{\star,\mathbbm{y}}_\alpha|\mathbbm{y}]\}\, ,\\
&=\textrm{E}_\mathbbm{y}\{\textrm{P}[\mathbbm{x}\in\mathcal{C}^{\mathbbm{y}}_\alpha|\mathbbm{y}]\} -(1-\alpha)\, ,
\end{split}
\end{equation}
which we will be able to approximate by Monte Carlo estimation techniques in \Cref{ssec:computation_ell}. 

When $\ell(\mathcal{C}^{\mathbbm{y}}_\alpha, \mathcal{C}^{\star,\mathbbm{y}}_\alpha)>0$, we say that $\mathcal{C}^{{y}}_\alpha$ is \emph{conservative}, as it overestimates uncertainty about $(\mathbbm{x}|\mathbbm{y}=y)$ on average, whereas for $\ell(\mathcal{C}^{\mathbbm{y}}_\alpha, \mathcal{C}^{\star,\mathbbm{y}}_\alpha)<0$ we say that $\mathcal{C}^{{y}}_\alpha$ is \emph{overconfident} or anti-conservative. When using \eqref{loss}, we pay particular attention to the range $\alpha \in (0,0.1)$ commonly encountered in UQ analyses. Moreover, while we prefer models for which the score \eqref{loss} is small in absolute value, we pay attention to the sign of \eqref{loss} and would generally argue in favour of models that are slightly conservative rather than overconfident. 

Note that \eqref{loss} is designed to probe the accuracy of UQ results in a manner that captures the repeated use of an imaging procedure: each application involves a new image $x^\star$ and a new observation $y$. In addition, \eqref{loss} is also easy to interpret and to compute numerically. However, because \eqref{loss} is only sensitive to the average error over the joint distribution of $(\mathbbm{x},\mathbbm{y})$, a method can in principle deliver very poor uncertainty results for some or all values of $y$ and still achieve a low value of $\ell(\mathcal{C}^{\mathbbm{y}}_\alpha, \mathcal{C}^{\star,\mathbbm{y}}_\alpha)$ if these errors average out over the space $\mathbb{Y}$. This limitation of \eqref{loss} is arguably not a major issue for imaging sciences yet, where reliable UQ is still challenging and currently available methods often markedly overestimate or underestimate uncertainty for a given $\alpha$. 

It is worth emphasising at this point that \eqref{loss} seeks to provide a mechanism for probing the quality of a Bayesian imaging techniques and their capacity to support UQ tasks, it does not seek to define the quality of the techniques. We discourage practitioners from seeking to directly optimise \eqref{loss} prior to its evaluation for a specific target value of $\alpha$. After the computation of \eqref{loss} has taken place, one can recalibrate a credible or confidence regions to achieve $\ell(\mathcal{C}^{\mathbbm{y}}_\alpha, \mathcal{C}^{\star,\mathbbm{y}}_\alpha) \approx 0$ for a target value of $\alpha$. This can be achieved, for example, by using conformal prediction techniques that use a calibration dataset \cite{Angelopoulos2023}. However, despite being operationally useful, applying a correction step will not improve in any manner the Bayesian model under scrutiny.

\subsection{Computation of \texorpdfstring{$\ell(\mathcal{C}^{\mathbbm{y}}_\alpha, \mathcal{C}^{\star,\mathbbm{y}}_\alpha)$}{the proposed error measure}}\label{ssec:computation_ell}
Computing $\ell(\mathcal{C}^{\mathbbm{y}}_\alpha, \mathcal{C}^{\star,\mathbbm{y}}_\alpha)$ exactly is not possible in practice because \emph{nature}'s distributions are complex and unknown. Therefore, we propose to use the following simple Monte Carlo estimator for \eqref{loss} summarised in \Cref{Algo:Proposed_Algorithm}, which relies on test dataset $\{x_i\}_{i=1}^M$ representing \emph{nature}'s marginal distribution for $\mathbbm{x}$.  
\begin{algorithm}[H]
            \caption{Proposed Monte Carlo algorithm to estimate $\ell(\mathcal{C}^{{y}}_\alpha, \mathcal{C}^{\star,{y}}_\alpha)$}
        \begin{algorithmic}
        \REQUIRE Test set $\{x_i\}_{i=1}^M$, Bayesian posterior density under scrutiny $p(x|y)$, level $\alpha \in (0,1)$.
                \FOR{$n = 1$ to $N$}
                \STATE 1. Draw a realisation from $\mathbbm{x}$ by randomly selecting an element $x_i$ of the test set $\{x_i\}_{i=1}^M$.
                \STATE 2. Draw a realization $y_i$ from $(\mathbbm{y}|\mathbbm{x}=x)$ by applying the forward observation model (or from the test set if pairs $\{x_i,y_i\}_{i=1}^M$ are available).
                \STATE 3. Using the Bayesian imaging model under scrutiny, compute a credible region that reportedly accumulates $1-\alpha$ posterior probability mass. For example, the region 
$$
\mathcal{C}^y_\alpha = \{x : p(x|y) \geq \gamma_\alpha\}
$$
where $\gamma_\alpha$ is set such that $\int_{\mathcal{C}^y_\alpha}p(x|y) \textrm{d}x = 1-\alpha$.
                \STATE 4. If $x_i \in \mathcal{C}^y_\alpha$, set $r_n = 0$. Otherwise, set $r_n = 1$.
                \ENDFOR
                \RETURN $\hat{\ell}(\mathcal{C}^{\mathbbm{y}}_\alpha, \mathcal{C}^{\star,\mathbbm{y}}_\alpha) = \alpha - \sum_{i=n}^{N} r_n/N$.
            \end{algorithmic}\label{Algo:Proposed_Algorithm}
\end{algorithm}

Note the high posterior density (HPD) region $
\mathcal{C}^y_\alpha = \{x : p(x|y) \geq \gamma_\alpha\}
$ can be easily reformulated in terms of the statistic $U_y(x) = \log p(y|x) + \log p(x)$ when working directly with $p(x|y)$ is not possible (e.g., because of an unknown normalisation factor). Similarly, one can implement \Cref{Algo:Proposed_Algorithm} with a different credible region, such as an $\ell_2$-ball centred on an estimate of the posterior mean, with radius adjusted to accumulate $1-\alpha$ posterior probability mass. This will lead to a region with larger volume, but does not affect in any manner the validity of the approach.

If the test set includes pairs $\{x_i,y_i\}_{i=1}^M$, then \Cref{Algo:Proposed_Algorithm} quantifies the misspecification error stemming from both the assumed observation model as well as from the assumed prior distribution. Otherwise, if $y_i$ in step 2 is a synthetic observation drawn from the considered forward observation model, then the likelihood function is assumed to be exact (unless the likelihood function is an approximation of the forward model). This latter case is useful for quantifying the impact of the misspecification of the prior on the credible regions $\mathcal{C}^y_\alpha$. This can provide an insight into the advantages and drawbacks of different prior elicitation strategies from the perspective of UQ, complementing the information that is already available from standard metrics  based on image estimation accuracy, such as the estimation PSNR.

\section{Evaluation of established and recent Bayesian imaging techniques}\label{sec:experiments}
\subsection{Experiment set up}
We now illustrate the proposed framework by applying it to the evaluation of a range of Bayesian imaging techniques in the context of a non-blind image deconvolution problem. For this evaluation, we use a standard linear observation model
\[ y = H x^\star + w,\]
where $x^\star$ and $y$ have size $96\times 96$ pixels\footnote{Assessing UQ accuracy requires a large number of experiment repetitions, hence the choice of small images.}, $H$ implements the discrete convolution with an uniform blur kernel of size $3\times 3$ pixels and $w$ is a realisation of white additive Gaussian noise, with variance set such such that the \emph{blurred signal-to-noise ratio} (BSNR) is 30dB. We have chosen this canonical image restoration problem as test case because it provides an indication of how Bayesian imaging methods will perform in an inverse problem that is moderately challenging. This problem is sufficiently ill-posed so that there is significant uncertainty about its solution, yet it can be tackled with a wide range of Bayesian imaging strategies without the need for extensive fine-tuning or unfeasibly long computing times. 

\begin{figure}
    \centering
    \adjustbox{width = 0.7\paperwidth}{\includegraphics{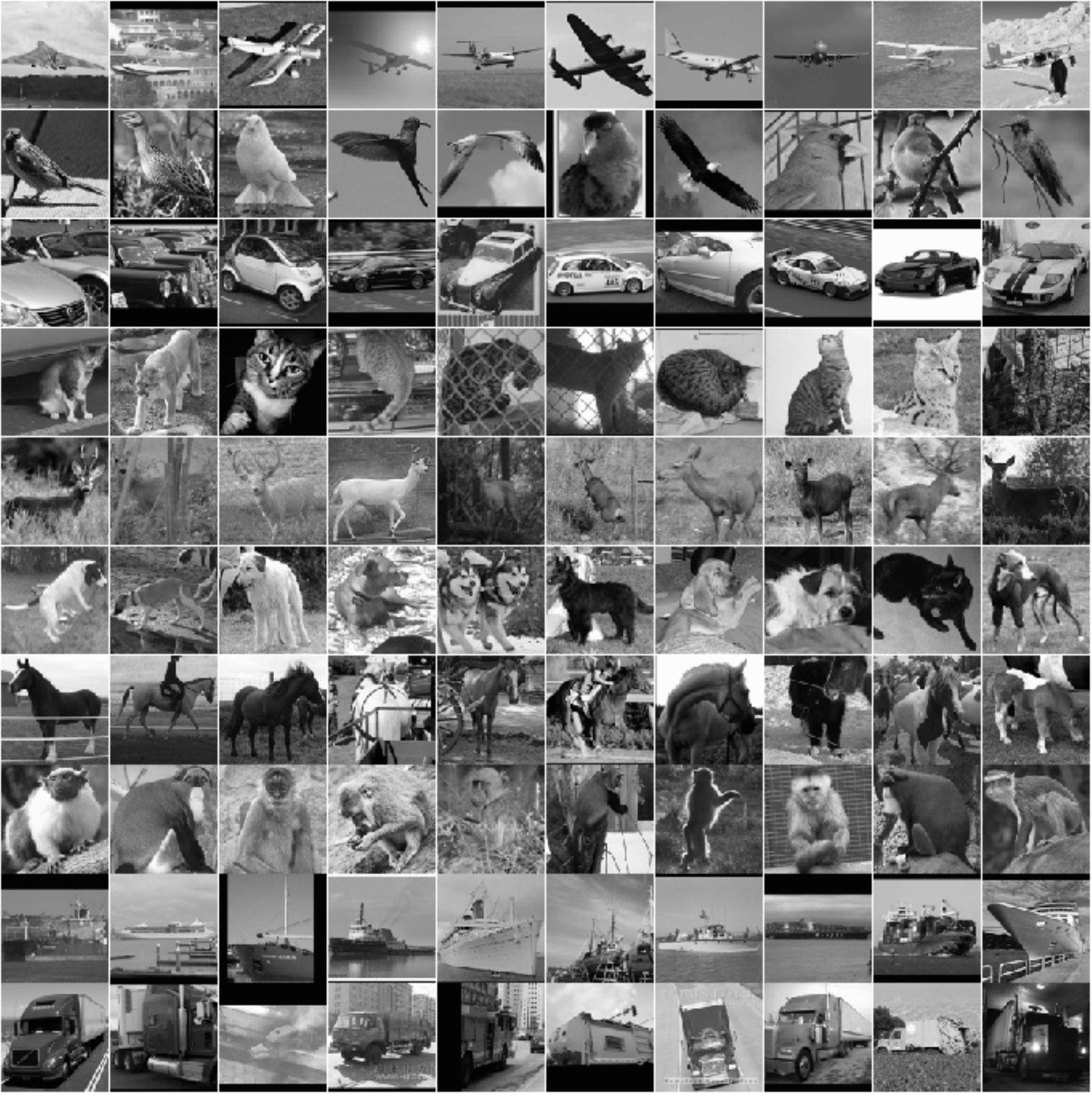}}
    \caption{Examples of images from the STL-10 dataset}
    \label{fig:stl10examples}
\end{figure}

As test dataset $\{x^\prime_i\}_{i=1}^N$, we use the STL-10 dataset \cite{coates2011}. The dataset consists of natural images of size $96\times 96$ pixels of airplanes, birds, cars, cats, deer, dogs, horses, monkeys, ships and trucks, see \Cref{fig:stl10examples}. Again, this is a relatively simple dataset that can be tackled straightforwardly with a wide range Bayesian imaging techniques. We use the greyscale version of the STL-10 dataset for classical image priors defined on scalar-valued pixels; otherwise we use the colour version of the STL-10 dataset.

With regards to the Bayesian imaging techniques considered, we have chosen to include in our experiment the following five techniques, which are representative of some of the main Bayesian strategies and trends for solving imaging inverse problems over the past two decades:

\paragraph{A hierarchical Bayesian model with a Gaussian Markov random field prior} We consider the hierarchical Bayesian method proposed in \cite{Orieux2010} which relies on a stationary Gaussian Markov random field prior with a circulant covariance matrix that promotes solutions with a low-pass power spectrum. The model also includes gamma conjugate hyperpriors to automatically adjust the scale parameters of the model. Inference is performed efficiently by using a Gibbs sampler that carefully exploits conjugacy properties and operates on the Fourier domain. This approach is representative of a wide class of Bayesian imaging strategies that originated with the seminal works \cite{Geman1984,green1990bayesian,besag1986statistical} four decades ago and remains widely used to date. A total of 20,000 samples were drawn for each image. As an estimate of the credible region, we use an $\ell_2$ ball centred on the posterior mean, with radius adjusted to accumulate $1-\alpha$ posterior probability mass, which is calculated from the samples drawn from the posterior distribution.

\paragraph{An empirical Bayesian model with a total-variation prior} This method relies on the (improper) total-variation (TV) prior $p(x|\lambda) \propto \exp\{-\lambda \mathrm{TV}(x)\}$, with $\mathrm{TV}(\x)=\sum_{i}\sqrt{(\Delta_{i}^{h}\x)^{2}+(\Delta_{i}^{v}\x)^{2}}$ denoting the isotropic TV pseudo-norm, where $\Delta_{i}^{v}$
and $\Delta_{i}^{h}$ denote horizontal and vertical first-order pixel
difference operators, and where $\lambda > 0$ is a regularisation parameter. Adopting an empirical Bayesian approach, $\lambda$ is automatically adjusted from $y$ by maximum marginal likelihood estimation. Leveraging the convexity of $x \mapsto \mathrm{TV}(x)$, one can perform empirical Bayesian inference for this model efficiently by using proximal Langevin Markov chain Monte Carlo and stochastic approximation proximal gradient schemes \cite{vidal:et:al:2019a}. This method is representative of a wide range of strategies based on non-smooth convex regularisation. Crucially, for such models, maximum-a-posteriori estimation is a convex optimisation problem that can be efficiently solved, even in very large settings, by using proximal splitting algorithms \cite{Parikh2014}. The Markov chain was warmed up and 20,000 samples were drawn from the posterior distribution $p(x|y)$.  We calculated the statistic $U_y(x) = \log p(y|x) + \log p(x)$ for each sample (up to an additive constant) and the HPD region by finding the quantile of this statistic such that $(1-\alpha)$\% of the samples lie within the credible region.

\paragraph{A Bayesian model with a log-concave data-driven prior} This method improves on the previous two methods by using a prior that is based on a convex ridge regulariser (CRR) \cite{goujon2023}, learnt from training data subject to a convexity constraint. The resulting log-concave prior is:
\[ p(x|\lambda) = \exp( - \lambda \sum_{i}^L \psi_i(w_{i}^{T}x)),\]
where each $\psi_i : \mathbb{R} \rightarrow \mathbb{R}$ is a learnt piece-wise quadratic (convex) activation function and $w_i$ is a learned filter. Following \cite{goujon2023},  we use $L = 32$ and learn the parameters of $\{\psi_i,w_i\}_{i=1}^L$ by using 238,000 grey-level patches of size $40 \times 40$ created from the BSD500 dataset. There are in the order of $5\,000$ parameters. The hyper-parameters $\lambda$ fixed by cross-validation.

Again, leveraging the underlying convexity, we perform maximum-a-posteriori estimation by optimisation, and other inferences by using a Langevin Markov chain Monte Carlo schemes. For each image, 40,000 samples, calculate the potential $U_y(x) = \log p(y|x) + \log p(x)$ for each sample (up to an additive constant), and the HPD region by finding the quantile of this statistic such that $(1-\alpha)$\% of the samples lie within the credible region.

\paragraph{ A Bayesian model with a plug-and-play prior represented by an image denoising algorithm} We now consider a state-of-the-art Bayesian imaging technique based on a plug-and-play (PnP) Langevin sampling scheme \cite{laumont_bayesian_2022}. Accordingly, we leverage Tweedie's identity \cite{efron_2011} and an image denoising algorithm $D_\varepsilon$ designed to remove additive Gaussian noise of variance $\varepsilon$ in order to construct the following approximation of the score function $x \mapsto  \nabla\log p^\star (x)$: 
 \[
 \nabla\log p^\star(x) \approx (D_{\varepsilon}(x)-x)/\varepsilon,\]
 where we recall that $p^\star$ denotes \emph{nature}'s prior to solve this Bayesian inference problem. This approximation can then be ``plugged'' into gradient-based Monte Carlo algorithms, such as the unadjusted Langevin algorithm (ULA), to perform approximate Bayesian computation w.r.t. \emph{nature}'s posterior $p^\star(x|y)$ \cite{laumont_bayesian_2022}. This strategy is particularly powerful when $D_\varepsilon$ is implemented by using a Lipschitz-regularised neural network trained for image denoising by using a large sample from the prior $p^\star$. In our experiments, we use the RealSN-DNN from \cite{Ryu2019PlugandPlayMP}. The network has 558,016 parameters and trained with noise level $5.0$ trained with the BDS500 dataset \cite{martin2001bsd500} divided into $40\times 40 $ patches, using ADAM for 50 epochs, with a mini-batch size of 128. The network is trained to have a Lipschitz constant of approximately 1.
 
With this approach, the statistic $x \mapsto \log p(x,y)$ is not available, hence the computation of the HPD region $\mathcal{C}^\star_\alpha$ is not tractable. Instead, we use an $\ell_2$ ball centred on the posterior mean, with radius adjusted to accumulate $1-\alpha$ posterior probability mass (as measured from the Monte Carlo samples produced by the PnP-ULA method \cite{laumont_bayesian_2022}). Note that this does not introduce any additional error, the resulting region is larger in volume but should nevertheless have the correct coverage. A total of 50,000 samples were drawn from the posterior distribution using PnP-ULA, and these samples were use to estimate the credible region.


\paragraph{A Denoising Diffusion Restoration Model (DDRM)}
Score-based denoising diffusion models have recently emerged as a powerful strategy to solve image restoration problems \cite{chung2022diffusion, zhu2023denoising, kawar2022denoising, laroche2024fast}. Here we consider the Denoising Diffusion Restoration Model (DDRM) \cite{kawar2022denoising}, which relies on the foundational score model with 280M parameters, trained by weighted denoising score-matching on the ImageNet dataset \cite{ho2020denoising}. Unlike previous approaches that rely on a Markov chain Monte Carlo strategy, denoising diffusion models construct a stochastic transport map between a standard normal random variable and the posterior distribution of interest. In practice, the implementation of this requires some careful approximations (DDRM leverages a highly efficient denoising diffusion implicit modelling (DDIM) approach specialised for image restoration problems with a Gaussian likelihood function \cite{kawar2022denoising}). In our experiments, for each image of the test dataset, we run 100 repetitions of DDRM to obtain 100 independent Monte Carlo samples for each image (we implement DDRM with 100 DDIM steps). Again, we use the obtained Monte Carlo samples to compute the posterior mean and an $\ell_2$ ball, centred on the mean, with radius adjusted to accumulate approximately $(1-\alpha)$ posterior probability mass. We use the implementation provided \cite{kawar2022denoising} and adjust hyper-parameters as recommended in \cite{kawar2022denoising}.

\subsection{Experimental results}
\begin{figure}[htbp]
    \centering
    \includegraphics[width=12cm]{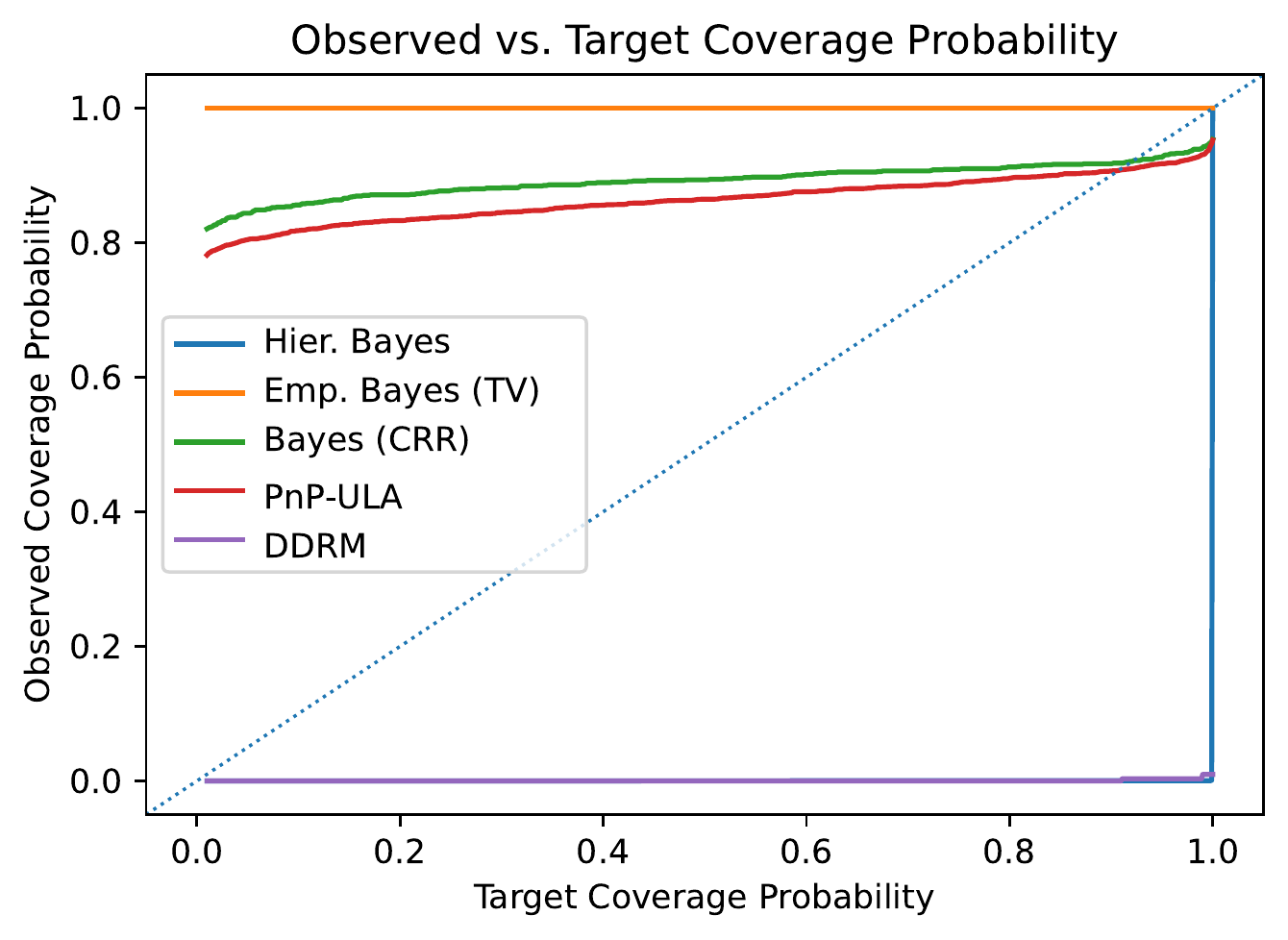}
  \caption{Observed empirical coverage probabilities vs target coverage probabilities $(1-\alpha)$ for each of the considered Bayesian imaging techniques}
  \label{fig:observed_vs_target_probs}
\end{figure}

We apply each of the five Bayesian imaging techniques mentioned above to the considered image deblurring problem. To measure the accuracy of the produced UQ results, we use \Cref{Algo:Proposed_Algorithm} to compute the discrepancy between the desired and the empirically observed $\alpha$, as measured by $\hat{\ell}(\mathcal{C}^{\mathbbm{y}}_\alpha, \mathcal{C}^{\star,\mathbbm{y}}_\alpha)$, for a fine grid of values of $\alpha$ ranging from $0.01$ to $0.99$. To ensure that probabilities are computed reliably, we use a large number of repetitions (i.e., we set the sample size in \Cref{Algo:Proposed_Algorithm} to $N = 2,500$). In addition, we also compute the average image reconstruction accuracy for each technique, as measured by peak-signal-to-noise-ration (PSNR), as well as the average computing time  per image (in minutes) to provide an indication of the relative computational cost of these different techniques. The results of this experiment are summarised in \Cref{tab:coverage} and \Cref{fig:observed_vs_target_probs}. Because of the large number of repetitions, the experiment required in the order of $1,100$ GPU hours (45 days) of computing time. 

We observe in \Cref{tab:coverage} that the average PSNR increases dramatically as we progress from classical to state-of-the-art Bayesian imaging techniques, with the denoising diffusion DDRM method achieving a remarkable $13$dB improvement on PSNR performance w.r.t. the hierarchical Bayesian method, at an expense of a $20\times$ increase in computing time. This important improvement captures the progress in estimation accuracy that has been achieved over the past few decades by harnessing ever more advanced probabilistic modelling strategies and large-scale machine learning techniques. Also note that there is a close agreement between the estimation performance of the methods and their chronological development, as expected.

Furthermore, we also observe in \Cref{tab:coverage} that the denoising diffusion DDRM method and the hierarchical Bayesian method perform extremely poorly in terms of UQ performance, with both methods producing highly overconfident credible regions that almost never contain the true solution. In the case of the hierarchical Bayesian approach, this poor performance is a consequence of the model being designed in a manner that favours certain algebraic (conjugacy) properties that are amenable to fast computation by traditional Gibbs sampling. However, in the case of the state-of-the-art DDRM method, the poor UQ results are arguably the consequence of models and algorithms that have been selected and optimised to achieve the best estimation performance in a computationally efficient manner, at the expense of other aspects of the posterior distribution. Interestingly, the fact that DDRM is highly overconfident also suggest that metrics form the generative modelling literature, such as the Frechet inception distance (FID), are not reliable proxies for posterior sampling accuracy.

Moreover, from \Cref{tab:coverage}, we would argue that the empirical Bayesian TV model is more reliable for UQ tasks than the state-of-the-art DDRM, as it is also inaccurate but produces credible regions that are conservative and almost surely contain the true solution. From the considered Bayesian methods, the two data-driven Langevin strategies achieve the most accurate credible regions, especially in the range where the target probability $(1-\alpha)$ is of the order of $80\% - 99\%$, as commonly encountered in UQ tasks. However, the UQ results delivered by the PnP-ULA method (with its RealSN-DNN network, $500k$ parameters) are very close to the results obtained with the simpler Bayesian CRR model, which has a much smaller network and a log-concavity constraint. This suggests that the architecture and training choices involved in the RealSN-DNN network, as well as the tuning of PnP-ULA, should be reconsidered.

Lastly, note that the results from \Cref{tab:coverage} and \Cref{fig:observed_vs_target_probs} are in close agreement with the UQ comparisons reported recently in Tachella and Pereyra in \cite{Tachella24} for compressive sensing, inpainting, and tomographic reconstruction with the DIV2K and LIDC datasets. This suggests that the conclusions from our experiment are likely to hold for other Gaussian linear problems.

  

\begin{table}[htbp]
    \centering
    \adjustbox{max width = \textwidth}{
    \begin{tabular}{@{}l*{7}{r}cr@{}}
\toprule 
 & \multicolumn{7}{c}{\textbf{Target Probability} $(1-\alpha)$} & \textbf{PSNR}  & \textbf{Time}\\
\cmidrule{2-8} 
 & $80.0$\,\% & $85.0$\,\% & $90.0$\,\% & $95.0$\,\% & $97.5$\,\% & $99.0$\,\% & $99.9$\,\%  & (dB)  & (mins)\\
\midrule 
\textbf{Hier. Bayes}  & $0.0$\,\%  & $0.0$\,\%  & $0.0$\,\%  & $0.0$\,\%  & $0.0$\,\%  & $0.0$\,\%  & $0.08$\,\%  & $30.1\pm2.0$  & 0.6\\ 
\textbf{Emp. Bayes (TV)}  & $100.0$\,\% & $100.0$\,\% & $100.0$\,\% & $100.0$\,\% & $100.0$\,\% & $100.0$\,\% & $100.0$\,\%  & $31.7\pm2.5$  & 1.0\\ \hdashline
\textbf{Bayes (CRR)}  & $91.2$\,\%  & $91.6$\,\%  & $91.7$\,\%  & $92.7$\,\%  & $93.4$\,\%  & $94.1$\,\%  & $95.0$\,\%  & $32.0\pm2.1$  & 5.3\\
\textbf{PnP-ULA} & $89.6$\,\% & $90.1$\,\% & $90.7$\,\% & $91.7$\,\% & $92.3$\,\% & $93.1$\,\% & $94.7$\,\%  & $33.0\pm2.5$  & 5.1\\
\textbf{DDRM}  & $0.0$\,\%  & $0.0$\,\%  & $0.0$\,\%  & $0.0$\,\%  & $0.5$\,\%  & $0.9$\,\%  & $1.8$\,\%  & $43.0\pm2.9$  & 13.5\\
\bottomrule 
\end{tabular}
}
    \caption{Coverage results for a $30$ dB SNR noise level: the target coverage probability $(1-\alpha)$ versus the observed coverage probability for each prior. The Peak Signal-to-Noise Ratio (PSNR) is given in decibels. Average computation times per each image are given in minutes. The horizontal dotted line divides the table into assumption (top) based versus data-driven priors.}
    \label{tab:coverage}
\end{table}

\section{Conclusion}\label{sec:conclusion}
This paper considered the fundamental questions of how to interpret the probabilities reported by a Bayesian imaging method and how to assess their accuracy from a frequentist viewpoint. These questions play a central role in uncertainty quantification and therefore in advancing imaging sciences for quantitative and scientific applications. We presented a novel Monte Carlo method for assessing accuracy of probabilities reported by Bayesian imaging techniques. We then leveraged the proposed methodology to assess five Bayesian imaging techniques that are representative of the main Bayesian imaging strategies of the past two decades. Our empirical findings suggest that reliable Bayesian UQ remains challenging, despite the sustained and significant progress in statistical image modelling of the past decades and the adoption of image priors based on machine learning. Some modern Bayesian imaging strategies can support some forms of UQ, but this needs to be checked carefully, as strategies that deliver excellent image estimation accuracy can still produce poor posterior distributions and unreliable UQ results. Also, metrics for generative modelling accuracy, such as the Frechet inception distance, are not reliable proxies for UQ accuracy.

To the best of our knowledge, the most accurate imaging UQ technique currently available is the equivariant bootstrap method of Tachella and Pereyra \cite{Tachella24}, which is not a Bayesian method. In fact, the method of \cite{Tachella24} still outperforms all Bayesian methods even in the unfavourable case where these methods use priors trained in a supervised manner and the equivariant bootstrap method is trained in a fully unsupervised way, directly from measurement data. This suggests that, with the appropriate design, future Bayesian imaging techniques should also be able to deliver accurate UQ results, even in cases where there is no ground truth data available. We hope that the availability of new methodology for evaluating Bayesian imaging techniques will help guide the development of statistical image models and lead to techniques that deliver accurate image estimates and Bayesian UQ inferences.

\section{Acknowledgments}
We are grateful to Andres Almansa, Gavin Gibson, and Julian Tachella for useful discussion.

\bibliographystyle{siamplain}
\bibliography{filtered_output3.bib}

\begin{thebibliography}{10}

\bibitem{salsa2010fast}
{\sc M.~V. Afonso, J.~M. Bioucas-Dias, and M.~A. Figueiredo}, {\em {Fast image recovery using variable splitting and constrained optimization}}, IEEE Transactions on Image Processing, 19 (2010), pp.~2345--2356, \url{https://doi.org/10.1109/tip.2010.2047910}.

\bibitem{Angelopoulos2023}
{\sc A.~N. Angelopoulos and S.~Bates}, {\em Conformal prediction: A gentle introduction}, Foundations and Trends in Machine Learning, 16 (2023), pp.~494--591, \url{https://doi.org/10.1561/2200000101}.

\bibitem{Bardsley2018-wx}
{\sc J.~M. Bardsley}, {\em Computational uncertainty quantification for inverse problems}, Computational Science and Engineering, Society for Industrial \& Applied Mathematics, New York, NY, Sept. 2018, \url{https://doi.org/10.1137/1.9781611975383.ch1}.

\bibitem{Bernardo1994}
{\sc J.~Bernardo and A.~Smith}, {\em {Bayesian Theory}}, Wiley, New York, 1994, \url{https://doi.org/10.1002/9780470316870}.

\bibitem{besag1986statistical}
{\sc J.~Besag}, {\em On the statistical analysis of dirty pictures}, Journal of the Royal Statistical Society. Series B (Methodological), 48 (1986), pp.~259--302, \url{https://doi.org/10.1111/j.2517-6161.1986.tb01412.x}.

\bibitem{buades_review_2005}
{\sc A.~Buades, B.~Coll, and J.~M. Morel}, {\em A {Review} of {Image} {Denoising} {Algorithms}, with a {New} {One}}, Multiscale Modeling \& Simulation, 4 (2005), pp.~490--530, \url{https://doi.org/10.1137/040616024}.

\bibitem{chambolle2016introduction}
{\sc A.~Chambolle and T.~Pock}, {\em An introduction to continuous optimization for imaging}, Acta Numerica, 25 (2016), pp.~161--319, \url{https://doi.org/10.1017/S096249291600009X}.

\bibitem{chatterjee2012patch}
{\sc P.~Chatterjee and P.~Milanfar}, {\em Patch-based near-optimal image denoising}, IEEE Transactions on Image Processing, 21 (2012), pp.~1635--1649, \url{https://doi.org/10.1109/TIP.2011.2172799}.

\bibitem{chen2023equivariant}
{\sc D.~Chen, M.~Davies, M.~J. Ehrhardt, C.-B. Schönlieb, F.~Sherry, and J.~Tachella}, {\em Imaging with equivariant deep learning: From unrolled network design to fully unsupervised learning}, IEEE Signal Processing Magazine, 40 (2023), pp.~134--147, \url{https://doi.org/10.1109/MSP.2022.3205430}.

\bibitem{chung2022diffusion}
{\sc H.~Chung, J.~Kim, M.~T. Mccann, M.~L. Klasky, and J.~C. Ye}, {\em Diffusion posterior sampling for general noisy inverse problems}, in The Eleventh International Conference on Learning Representations, 2023, \url{https://arxiv.org/abs/2209.14687}.

\bibitem{coates2011}
{\sc A.~Coates, A.~Ng, and H.~Lee}, {\em An analysis of single-layer networks in unsupervised feature learning}, Journal of Machine Learning Research - Proceedings Track, 15 (2011), pp.~215--223, \url{https://proceedings.mlr.press/v15/coates11a.html}.

\bibitem{Durmus2018}
{\sc A.~Durmus, E.~Moulines, and M.~Pereyra}, {\em Efficient {Bayesian} computation by {Proximal Markov Chain Monte Carlo}: When {Langevin} meets {Moreau}}, SIAM Journal on Imaging Sciences, 11 (2018), p.~473–506, \url{https://doi.org/10.1137/16m1108340}, \url{http://dx.doi.org/10.1137/16M1108340}.

\bibitem{efron_2011}
{\sc B.~Efron}, {\em Tweedie's formula and selection bias}, Journal of the American Statistical Association, 106 (2011), pp.~1602--1614, \url{https://doi.org/10.1198/jasa.2011.tm11181}.

\bibitem{Geman1984}
{\sc S.~Geman and D.~Geman}, {\em {S}tochastic relaxation, {G}ibbs distributions, and the {B}ayesian restoration of images}, {IEEE} Transactions on Pattern Analysis and Machine Intelligence, 6 (1984), pp.~721--741, \url{https://doi.org/10.1109/TPAMI.1984.4767596}.

\bibitem{gonzalez2022solving}
{\sc M.~Gonz\'{a}lez, A.~Almansa, and P.~Tan}, {\em Solving inverse problems by joint posterior maximization with autoencoding prior}, SIAM Journal on Imaging Sciences, 15 (2022), pp.~822--859, \url{https://doi.org/10.1137/21M140225X}.

\bibitem{goujon2023}
{\sc A.~Goujon, S.~Neumayer, P.~Bohra, S.~Ducotterd, and M.~Unser}, {\em A neural-network-based convex regularizer for inverse problems}, IEEE Transactions on Computational Imaging,  (2023), pp.~1--15, \url{https://doi.org/10.1109/TCI.2023.3306100}.

\bibitem{green1990bayesian}
{\sc P.~Green}, {\em Bayesian reconstructions from emission tomography data using a modified {EM} algorithm}, IEEE Transactions on Medical Imaging, 9 (1990), pp.~84--93, \url{https://doi.org/10.1109/42.52985}.

\bibitem{ho2020denoising}
{\sc J.~Ho, A.~Jain, and P.~Abbeel}, {\em Denoising {Diffusion} {Probabilistic} {Models}}, in Advances in {Neural} {Information} {Processing} {Systems}, H.~Larochelle, M.~Ranzato, R.~Hadsell, M.~F. Balcan, and H.~Lin, eds., vol.~33, Curran Associates, Inc., 2020, pp.~6840--6851, \url{https://proceedings.neurips.cc/paper_files/paper/2020/file/4c5bcfec8584af0d967f1ab10179ca4b-Paper.pdf}.

\bibitem{holden2022bayesian}
{\sc M.~Holden, M.~Pereyra, and K.~C. Zygalakis}, {\em Bayesian imaging with data-driven priors encoded by neural networks}, SIAM Journal on Imaging Sciences, 15 (2022), pp.~892--924, \url{https://doi.org/10.1137/21M1406313}.

\bibitem{kaipio2006statistical}
{\sc J.~Kaipio and E.~Somersalo}, {\em {Statistical and computational inverse problems}}, vol.~160, Springer Science \& Business Media, 2006, \url{https://doi.org/10.1007/b138659}.

\bibitem{kawar2022denoising}
{\sc B.~Kawar, M.~Elad, S.~Ermon, and J.~Song}, {\em Denoising {Diffusion} {Restoration} {Models}}, in Advances in {Neural} {Information} {Processing} {Systems}, S.~Koyejo, S.~Mohamed, A.~Agarwal, D.~Belgrave, K.~Cho, and A.~Oh, eds., vol.~35, Curran Associates, Inc., 2022, pp.~23593--23606, \url{https://proceedings.neurips.cc/paper_files/paper/2022/file/95504595b6169131b6ed6cd72eb05616-Paper-Conference.pdf}.

\bibitem{laroche2024fast}
{\sc C.~Laroche, A.~Almansa, and E.~Coupete}, {\em {Fast diffusion EM: A diffusion model for blind inverse problems with application to deconvolution}}, in Proceedings of the IEEE/CVF Winter Conference on Applications of Computer Vision, 2024, pp.~5271--5281, \url{https://doi.org/10.1109/wacv57701.2024.00519}.

\bibitem{laumont_bayesian_2022}
{\sc R.~Laumont, V.~D. Bortoli, A.~Almansa, J.~Delon, A.~Durmus, and M.~Pereyra}, {\em Bayesian {Imaging} {Using} {Plug} \& {Play} {Priors}: {When} {Langevin} {Meets} {Tweedie}}, SIAM Journal on Imaging Sciences, 15 (2022), pp.~701--737, \url{https://doi.org/10.1137/21M1406349}.

\bibitem{liaudat2023_3}
{\sc T.~I. Liaudat, M.~Mars, M.~A. Price, M.~Pereyra, M.~M. Betcke, and J.~D. McEwen}, {\em Scalable {Bayesian} uncertainty quantification with data-driven priors for radio interferometric imaging}, arXiv preprint arXiv:2312.00125,  (2023), \url{https://doi.org/10.48550/arXiv.2312.00125}.

\bibitem{martin2001bsd500}
{\sc D.~Martin, C.~Fowlkes, D.~Tal, and J.~Malik}, {\em A database of human segmented natural images and its application to evaluating segmentation algorithms and measuring ecological statistics}, in Proceedings Eighth IEEE International Conference on Computer Vision. ICCV 2001, vol.~2, 2001, pp.~416--423 vol.2, \url{https://doi.org/10.1109/ICCV.2001.937655}.

\bibitem{Mukherjee23}
{\sc S.~Mukherjee, A.~Hauptmann, O.~{\"{O}}ktem, M.~Pereyra, and C.~Sch{\"{o}}nlieb}, {\em Learned reconstruction methods with convergence guarantees: {A} survey of concepts and applications}, {IEEE} Signal Process. Mag., 40 (2023), pp.~164--182, \url{https://doi.org/10.1109/MSP.2022.3207451}.

\bibitem{Ongie2020}
{\sc G.~Ongie, A.~Jalal, C.~A. Metzler, R.~G. Baraniuk, A.~G. Dimakis, and R.~Willett}, {\em Deep learning techniques for inverse problems in imaging}, IEEE Journal on Selected Areas in Information Theory, 1 (2020), pp.~39--56, \url{https://doi.org/10.1109/JSAIT.2020.2991563}.

\bibitem{Orieux2010}
{\sc F.~Orieux, J.-F. Giovannelli, and T.~Rodet}, {\em {Bayesian estimation of regularization and point spread function parameters for Wiener--Hunt deconvolution}}, Journal of The Optical Society of America, 27 (2010), \url{https://doi.org/10.1364/josaa.27.001593}.

\bibitem{pan2013l0}
{\sc J.~Pan and Z.~Su}, {\em Fast $\ell ^{0}$-regularized kernel estimation for robust motion deblurring}, IEEE Signal Processing Letters, 20 (2013), pp.~841--844, \url{https://doi.org/10.1109/LSP.2013.2261986}.

\bibitem{Parikh2014}
{\sc N.~Parikh}, {\em Proximal algorithms}, Foundations and Trends{\textregistered} in Optimization, 1 (2014), p.~127–239, \url{https://doi.org/10.1561/2400000003}, \url{http://dx.doi.org/10.1561/2400000003}.

\bibitem{pereyra2016survey}
{\sc M.~Pereyra, P.~Schniter, E.~Chouzenoux, J.-C. Pesquet, J.-Y. Tourneret, A.~O. Hero, and S.~McLaughlin}, {\em {A survey of stochastic simulation and optimization methods in signal processing}}, IEEE Journal of Selected Topics in Signal Processing, 10 (2016), pp.~224--241, \url{https://doi.org/10.1109/JSTSP.2015.2496908}.

\bibitem{Tachella24}
{\sc M.~Pereyra and J.~Tachella}, {\em Equivariant bootstrapping for uncertainty quantification in imaging inverse problems}, in Proceedings of The 27th International Conference on Artificial Intelligence and Statistics, S.~Dasgupta, S.~Mandt, and Y.~Li, eds., vol.~238 of Proceedings of Machine Learning Research, PMLR, 02--04 May 2024, pp.~4141--4149, \url{https://proceedings.mlr.press/v238/pereyra24a.html}.

\bibitem{repetti2019scalable}
{\sc A.~Repetti, M.~Pereyra, and Y.~Wiaux}, {\em Scalable {Bayesian} uncertainty quantification in imaging inverse problems via convex optimization}, SIAM Journal on Imaging Sciences, 12 (2019), pp.~87--118, \url{https://doi.org/10.1137/18M1173629}.
\newblock Publisher: Society for Industrial and Applied Mathematics.

\bibitem{robert2007bayesian}
{\sc C.~Robert}, {\em The {Bayesian} Choice}, Springer {Texts} in {Statistics}, Springer New York, New York, NY, 2007, \url{https://doi.org/10.1007/0-387-71599-1}.

\bibitem{roininen2019hyperpriors}
{\sc L.~Roininen, M.~Girolami, S.~Lasanen, and M.~Markkanen}, {\em Hyperpriors for {Matérn} fields with applications in {Bayesian} inversion}, Inverse Problems and Imaging, 13 (2019), pp.~1--29, \url{https://doi.org/10.3934/ipi.2019001}.

\bibitem{Ryu2019PlugandPlayMP}
{\sc E.~K. Ryu, J.~Liu, S.~Wang, X.~Chen, Z.~Wang, and W.~Yin}, {\em {Plug-and-Play} methods provably converge with properly trained denoisers}, in International Conference on Machine Learning, 2019, \url{http://proceedings.mlr.press/v97/ryu19a.html}.

\bibitem{Sullivan2019-dr}
{\sc T.~J. Sullivan}, {\em Introduction to Uncertainty Quantification}, Texts in Applied Mathematics, Springer International Publishing, Cham, Switzerland, March 2019, \url{https://doi.org/10.1007/978-3-319-12385-1_1}.

\bibitem{vidal:et:al:2019a}
{\sc A.~F. Vidal, V.~De~Bortoli, M.~Pereyra, and D.~Alain}, {\em Maximum likelihood estimation of regularisation parameters in high-dimensional inverse problems: an empirical {Bayesian} approach. {Part I}: Methodology and experiments.}, SIAM Journal on Imaging Sciences,  (2020), \url{https://doi.org/10.1137/20m1339829}.

\bibitem{zhu2023denoising}
{\sc Y.~Zhu, K.~Zhang, J.~Liang, J.~Cao, B.~Wen, R.~Timofte, and L.~Van~Gool}, {\em {Denoising Diffusion Models} for {Plug-and-Play} image restoration}, in {Proceedings of the IEEE/CVF Conference on Computer Vision and Pattern Recognition}, 2023, pp.~1219--1229, \url{https://doi.org/10.1109/cvprw59228.2023.00129}.

\end{thebibliography}
\end{document}